\documentclass[manuscript]{aastex}	% For Apj manuscript Draft format
\usepackage{epsfig}			% For eps figures, old commands
\usepackage{amssymb}			% For useful mathematical symbols
\usepackage{color}			% For color text: \color command
\usepackage{url}			% For breaking URLs easily trough lines
\usepackage{amsmath}			% Provides various features to facilitate writing math formulas
\usepackage{rotating}			% For rotate any object of an arbitrary angle
\usepackage{float}			% For improveing the interface for defining floating objects
\usepackage{textcomp}			% Pro­vides many text sym­bols (such as baht, bul­let, copy­right,etc )
\usepackage{psfig}
\usepackage{dcolumn}
\usepackage{times}
\usepackage{tabularx}
\usepackage[english]{babel}
\usepackage{float}

\renewcommand{\deg}{\ensuremath{^\circ}}
\newcommand{\unit}[1]{\ensuremath{\,\mathrm {#1}}}  % text in math mode

\shorttitle{Forward Modelling of Propagating Slow wave damping}
\shortauthors{S. Mandal et al.}

\begin{document}

%%%%%---------------Title----------------%%%%%

\title{Forward Modelling of Propagating Slow Waves in Coronal Loops and Their Frequency-Dependent Damping.}

\author{Sudip Mandal$^{1}$,
Norbert Magyar$^{2}$,
Ding Yuan$^{2}$,
Tom Van Doorsselaere$^{2}$,
Dipankar Banerjee$^{1,3}$}
 
\affil{$^{1}$Indian Institute of Astrophysics, Koramangala, Bangalore 560034, India. e-mail: {\color{blue}{sudip@iiap.res.in}}\\
$^{2}$ Center for mathematical Plasma Astrophysics, Department of Mathematics, KU Leuven, Celestijnenlaan 200B, bus 2400, 3001, Leuven, Belgium\\
$^{3}$ Center of Excellence in Space Sciences India, IISER Kolkata, Mohanpur 741246, West Bengal, India }

%%%---------------------Abstract-------------------------%%%%
\begin{abstract}
{ Propagating slow waves in coronal loops exhibit a damping which depends upon the frequency of the waves. In this study we aim to investigate the relationship of the damping length (L$_d$) with the frequency of the propagating wave. We present a 3-D coronal loop model with uniform density and temperature and investigate the frequency dependent damping mechanism for the four chosen wave periods. We include the thermal conduction to damp the waves as they propagate through the loop. The numerical model output has been forward modelled to generate synthetic images of SDO/AIA 171 \r{A} and 193 \r{A} channels. The use of forward modelling, which incorporates the atomic emission properties into the intensity images, allows us to directly compare our results with the real observations. The results show that the damping lengths vary linearly with the periods. We also measure the contributions of the emission properties on the damping lengths by using density values from the simulation. In addition to that} we have also calculated the theoretical dependence of L$_d$ with wave periods and showed that it is consistent with the results we obtained from the numerical modelling and earlier observations.

\end{abstract}
% --- Sun: magnetic fields}

% \online{: animation, color figures}
%%%%%-----------------Introductory section-----------------%%%%%
\section{Introduction}

Propagating intensity disturbances (PDs), observed with SoHO/UVCS \citep{1997ApJ...491L.111O} and SoHO/EIT \citep{1998ApJ...501L.217D,berghmans1999active} and TRACE \citep{nightingale2000time,2000A&A...355L..23D}, have been interpreted as propagating slow-magnetoacoustic waves \citep{1999ApJ...514..441O,2000ApJ...533.1071O,2012ASPC..456...91W}. These PDs are omnipresent in the solar corona \citep{2012A&A...546A..50K} and they propagate with an apparent speed ranging from 50 to 150 km~s$^{-1}$ which is close to the sound speed in the corona \citep{2012SoPh..279..427K,2009ApJ...697.1674M,2009ApJ...706L..76M}. However recent spectroscopic analysis shows that these disturbances could well be upwardly propagating flows channeling through the loop systems \citep{2010ApJ...722.1013D,2011ApJ...727L..37T,2011ApJ...738...18T}. Using coherent line parameter oscillations and `Red-Blue' asymmetry analysis \citet{2009ApJ...701L...1D} and \citet{2011ApJ...738...18T} showed that these PDs are more like upflows rather than waves. But, \citet{2010ApJ...724L.194V} have shown that upward propagating slow waves generally have the tendency to enhance the blue wing of the emission line because of the in-phase behaviour of velocity and density perturbations.
\citet{2011ApJ...737L..43N} pointed out that the flows are dominant near the footpoints of the loops and their strength decreases as we go away leading to a dominant wave scenario. Using forward modelling \citet{2015SoPh..290..399D} calculated the spectroscopic signatures of waves and flows, but hardly found an observable to distinguish these two phenomena. Recently \citet{2015RAA....15.1832M} performed detailed analysis using simultaneous imaging (SDO/AIA) and spectroscopic (HINODE/EIS) data and concluded that with the current instrumental capabilities it is very difficult to decouple them from each other. 

Slow waves are often used for seismological studies \citep{2003A&A...404L...1K,2009A&A...503L..25W,2011ApJ...727L..32V,2012A&A...543A...9Y}. These waves also get damped as they propagate along the magnetic structures. Damping of slow waves has been studied intensively from observations as well as theoretical modelling \citep{2002ApJ...580L..85O,2003A&A...408..755D,2004A&A...415..705D,2004A&A...425..741D,
2005A&A...436..701S,2005A&A...437L..47V,2014ApJ...789..118K,2014A&A...568A..96G,2015arXiv150504475B}. Using boundary driven oscillations and including thermal conduction and viscosity as the damping mechanisms \citet{2003A&A...408..755D} found the thermal conduction to be the dominant mechanism for damping of propagating slow waves in typical coronal condition and these results match very well with the observed damping lengths from TRACE observations. The contributions of gravitational stratification, field line divergence \citep{2004A&A...415..705D} and the mode coupling \citep{2004A&A...425..741D} on the wave damping are negligible compared to the damping due to thermal conduction. The dependence of the damping lengths with the observed periods have also been studied in the past. Propagating slow waves with different periodicities are detected up to different heights in the solar atmosphere and thus it indicates a frequency dependent damping mechanism operating on these waves \citep{2012SoPh..281...67K}. Using observational data from SDO/AIA, \citet{2014ApJ...789..118K} investigated the frequency dependent damping of the propagating slow waves in on-disk sunspot loops as well as polar plume-interplume structures.
 In the low thermal conduction limit, the theory predicts a slope of 2 in the log-log plot of damping length versus the wave period but these authors found a small positive slope (less than 2) for the on-disk structures and even negative slopes for the polar structures. They concluded that the deviations of these slope values indicate some missing element in the damping theory and may point towards the existence (or dominance) of a different damping mechanism, other than thermal conduction, operating in these structures. Studying the slow waves in a polar coronal hole, \citet{2014A&A...568A..96G} found a frequency dependent damping behavior for which the lower-frequency waves can travel to greater heights whereas the higher-frequency waves are damped heavily. The author also found that the wave is getting damped heavily within the first 10 Mm and after that damping affects the waves slowly with height.

The forward modelling technique has been used previously to study various phenomena in the solar corona (\citet{2008SoPh..252..101D,2014ApJ...787L..22A,2015ApJ...806...81A,2015ApJ...807...98Y} and references within). 
\citet{2008SoPh..252..101D} studied the intensity perturbations using the forward modelling in a long coronal loop. They synthesized the SOHO/TRACE 171\r{A} and the  HINODE/EIS 195\r{A} line and found that the observed intensity perturbation need not necessarily follow the model density and temperature. The discrepancy comes because of nonlinear interaction between the density, ionization balance and the instrumental response function. This shows the necessity of the forward modelling when comparing the numerical model results with the real observation.

In this work we set out to explain the discrepancies between the observed and theoretical values of the slope in a log-log plot of the damping length versus period. We present a 3-D numerical loop model (Section \ref{Numerical Model}) with constant density and temperature along the loop length and implemented anisotropic thermal conduction as the damping mechanism. In our model we have not included the gravitational stratification and the flux tube divergence as they have no contributions towards the frequency dependent damping length (see Table 1 in \citet{2014ApJ...789..118K}). We have used forward modelling (Section \ref{FoMo}) to generate synthesized SDO/AIA images from our numerical model output. The forward modelling technique converts the model output parameter (e.g density) into observable (e.g intensity) with the use of the instrumental filter response function. Thus the synthesized images contain information about the MHD wave theory as well as the atomic emissions as would be observed by the SDO/AIA. Here we want to emphasize the fact that we use a 3-D model to fully exploit the advantage of forward modelling (which takes into account the effects due to LoS angle, column depth and pixel size \citep{2003A&A...397..765C}) to create synthesized images which allow us to compare our results immediately with the observation which was not possible with the previously mentioned 1-D loop models.The detailed analysis and results of the synthesized data is described in Section \ref{Analysis}. Analytical solutions from the related theory and the conclusions are presented in the subsequent sections.

%%%%%--------------------Main body-------------------------%%%%% 
%----------------------------1---------------------------------
\section{Numerical Model} \label{Numerical Model}

\begin{figure}
\centering
\includegraphics[angle=0,width=0.9\textwidth]{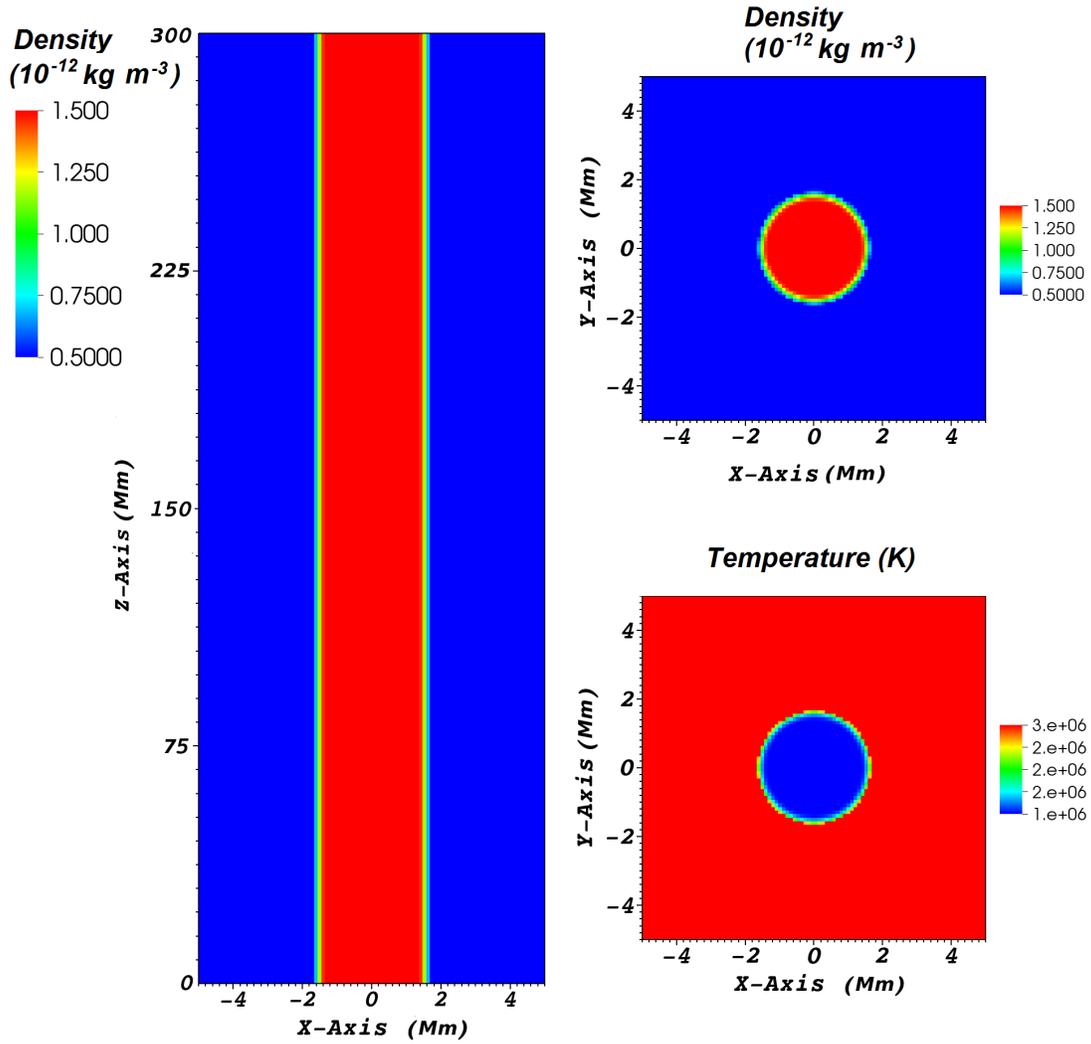}
\caption{Plot showing the initial condition: slice along the tube (left) and cross-section (right) showing the density (left, upper right) and temperature (lower right). The spatial dimension of the numerical domain is $10\ \mathrm{Mm} \times 10\ \mathrm{Mm} \times 300\ \mathrm{Mm}$ (for the multi-period runs).}
\label{model} 
\end{figure}

Our 3D numerical model consists of a straight, density enhanced flux tube embedded in a background plasma, the whole region being permeated by a uniform magnetic field, parallel to the flux tube (see Figure~\ref{model}). The density varies smoothly from the interior of the flux tube to the  background value, with an inhomogeneous layer (at the boundaries perpendicular to the loop length) of width $l \approx 0.1R$, for numerical stability. We neglect the effect of gravity and loop curvature. The values of relevant physical parameters are listed in Table~\ref{ICvalues}. 

%%%%%%%%%%%%%%%%%%%%%%%%%%%%%%%%%%%%%%%%%%%%%%%%%%%%%%%%%%%%%%%%

%%%%%%%%%%%%%%%%%%%%%%%%%%%%%%%%%%%%%%%%%%%%%%%%%%%%%%%%%%%%%%%%

In order to excite slow waves in the flux tube, we perturb the pressure at its footpoint near the boundary of the computational domain, acting only inside the flux tube and going to zero outside the loop diameter. This time-dependent pressure perturbation \( (p\prime)\) has the following form:

\begin{equation}
\dfrac{p'(t)}{p_0} = \mbox{constant} +\sum_{i=1}^n \alpha_i \sin\left(\frac{2 \pi t}{T_i}\right),
\end{equation}
where $p_0$ is the initial pressure, $n$ is the number of different single-period waves, $\alpha_i$ is the relative amplitude of the perturbation, and $T_i$ is the wave period. We drive the waves during the whole simulation time. Simulations were run with one ($n=1$) and four ($n=4$) periods. The relative amplitudes chosen for the multi-period driver are  $\alpha_1 = 0.2, \alpha_2=0.203, \alpha_3=0.206, \alpha_4=0.21$, while for the single-period driver, $\alpha_1 = 0.2$. The corresponding wave periods are $T_1 = 3\ \mathrm{min}, T_2 = 5\ \mathrm{min}, T_3 = 7\ \mathrm{min}, T_4 = 10\ \mathrm{min}$  and $T_1 = 3\ \mathrm{min}$.
 At this (bottom) boundary, we have reflective boundary conditions for the velocity, and zero-gradient for the other variables. At the top end of the flux tube, we apply open (or zero-gradient) boundary conditions, letting the waves leave the domain, though producing minor reflections. The other perpendicular boundaries are periodic. Note that the tube length for the multi-period runs is longer than that of the single-period runs, in order to accommodate three wavelengths of the longest period wave ($T_4$). This is necessary in order to get a reliable fit to the damped amplitudes in the analysis to follow.

The MHD problem is solved numerically, using \texttt{MPI-AMRVAC} \citep{2012JCoPh.231..718K,2014ApJS..214....4P} using HLL solver . The uniform grid consists of $96 \times 96 \times 600$ numerical cells, enough to eliminate the damping of the oscillations due to numerical dissipation, as test runs have shown.
We include anisotropic thermal conduction, i.e. along the magnetic field lines, with the Spitzer conductivity set to $\kappa = 10^{-6}\ T^{5/2}\ \mathrm{erg\ cm^{-1}\  s^{-1}\  K^{-1}}$, where $T$ is the temperature (see \citet{2011ApJ...727L..32V,2012ApJ...748L..26X}).

%%%-------------------------------------------------

  \begin{table}[h]
  \caption{\label{ICvalues}The values of relevant physical parameters used in the simulations.}
  	\centering
  	\begin{tabular}{lccc}
  		\hline
  		$Parameter$ & $Value$ \\ 
  		\hline
  		Tube length (multi-period run) ($L$)& 300 Mm \\
  		Tube length (single-period run) ($L$)& 100 Mm \\
  		Tube radius ($R$)& 1.5 Mm \\
  		Magnetic Field & 12.5 G \\
  		Tube density \((\rho_\mathrm{inside})\) & $1.5 \cdot 10^{-12}$ kg/m$^3$ \\
  		Density ratio ($\rho_\mathrm{inside}/ \rho_\mathrm{outside}$)& 3 \\
  		Tube temperature & 1.0 MK \\
  		Background plasma temperature & 3.0 MK \\
  		Plasma $\beta$ & 0.04 \\
  		\hline
  	\end{tabular}
  	\label{table}
  \end{table}
%%%-------------------------------------------------

\section{Synthesizing AIA observation} \label{FoMo}

We synthesize imaging observations in the SDO/AIA 171 \AA{} ($\approx0.6\unit{MK}$) and 193 \AA{} ($\approx ~1.2\unit{MK}$) bandpasses, within which the $1\unit{MK}$ loop could be prominently detected. We use the FoMo code\footnote{\url{https://wiki.esat.kuleuven.be/FoMo/FrontPage}} to perform forward modelling. It has been used for forward modelling by \citet{2013A&A...555A..74A,2014ApJ...787L..22A,2015ApJ...806...81A,2015ApJ...807...98Y}. The FoMo code uses the AIA temperature response function \citep{2011A&A...535A..46D,2012SoPh..275...41B} and performs numerical integration along selected LOS angles \citep[see details in][]{2015ApJ...807...98Y}. We obtain synthetic AIA observation along LOS angles at $30\deg$ and $90\deg$ (see Figure\ref{single_damping}) for both bandpasses. The output image data are coarsened to AIA pixel size ($0.6\arcsec$); while in the numerical LOS integrations of the emission, the discretization has a slightly better resolution than the numerical simulation. The time interval between two successive synthesized AIA images is 23 seconds.

\section{Analysis of the model output} \label{Analysis}

\subsection{Single-period driver}

\begin{figure}[H]
\centering
\includegraphics[angle=90,width=0.9\textwidth]{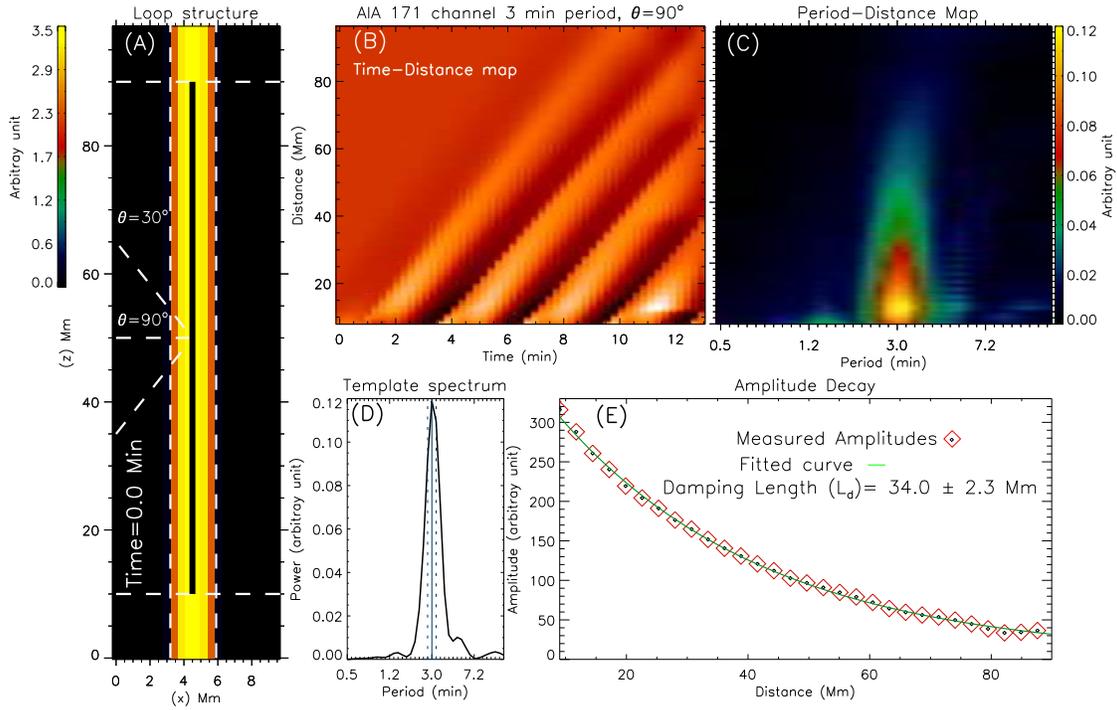}
\caption{$(A):$ The initial intensity snapshot for AIA 171~\r{A} channel for $\theta$= 90\textdegree. Two horizontal white dashed line indicate the region of the loop selected for the analysis. The LOS angles ($\theta$) are also marked in the panel $(B):$ Time-distance plot produced from AIA 171\r{A} image sequence by placing an artificial slit marked by a black vertical line in panel-A. $(C)$ and $(D):$ corresponding period-distance map and the template spectrum (made from bottom 5 pixels of panel-c) using the time-distance map. $(E):$ The amplitude decay plotted with the distance along the loop length. The damping length obtained from a fitted exponential function is printed on the plot.}
\label{single_damping} 
\end{figure}

 The initial intensity image (before switching on the driver) for the AIA 171~\r{A} channel is shown in panel-A of Figure \ref{single_damping}.

%%%%%%%%%%%%%%%%%%%%%%%%%%%%%%%%%%%%%%%%%%%%%%%%%%%%%%%%%%%%%%%%
%%%%%%%%%%%%%%%%%%%%%%%%%%%%%%%%%%%%%%%%%%%%%%%%%%%%%%%%%%%%%%%%

 In the analysis, we choose our region of interest along the length of the loop leaving $\approx$10 Mm, from each side of the loop-footpoints, to avoid any kind of boundary effects (shown as horizontal white dashed lines in panel-A). In this paper, we followed a similar approach as used by \citet{2014ApJ...789..118K}. To generate a slow wave with a period of 3 minutes, we implement a continuous driver of the same period situated at one of the loop footpoint. From the movie (movie~1, available online) we see that the waves, propagating from one footpoint to the other, are getting damped as they move along the length of the loop. This property is clearly visible in the time-distance map which we construct by placing an artificial slit of thickness 1 Mm along the length of the loop, shown as a black thick line in panel-A. To construct an enhanced time-distance map (panel-B), we have subtracted a 15 points (i.e time frames) running difference from the original map to remove the background.

To analyze the propagation of the power along the distance of the loop, we convert the time-distance map into a period-distance map. We use the wavelet transform \citep{1998BAMS...79...61T} on the time series at each spatial pixel on the time-distance map to obtain the period-distance map. As shown in panel-C of Figure \ref{single_damping}, we notice that the power is concentrated only around the 3 minute period (expected as the driver period is 3 minutes) and the fact that the power decreases as we move along the length of the loop. The power distribution as function of the period is plotted in panel-D. This template spectrum is obtained using the bottom 5 pixels of the period distance map. To obtain a trend of the power decrement, we follow the amplitude (which is the square root of power) along the 3 minute period (within the width, as shown in dotted lines in panel-D, obtained from a SolarSoft routine $gt\_peaks$.pro ) and plot them with the distance of the loop. We plot every $15^{th}$ point in the panel to avoid crowding of points. The amplitude decay is then fitted with an exponential function of the form \(A(y) = A_0 e^{-y/L_d}+C\) where $L_d$ is the damping length of that period. The obtained damping length is equal to 34.0$\pm $2.3  Mm.

\subsection{Multi-period Driver}

\begin{figure}[h]
\centering
\includegraphics[angle=90,width=0.8\textwidth]{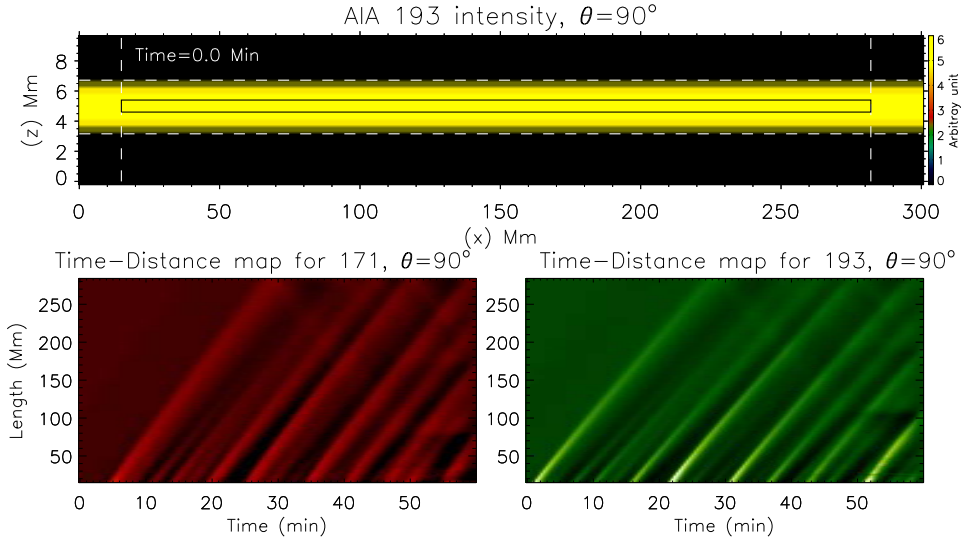}
\caption{$Top~panel:$ Initial intensity image for AIA 193\r{A} channel for $\theta$=90\textdegree. The Black rectangular box indicates the artificial slit used to create the time-distance maps. $Bottom~panel:$ Time-distance maps, for the multiperiod driver (3,5,7 and 10 minutes), are shown for AIA 171\r{A} and 193\r{A} channel respectively. The inclined ridges indicate the propagating slow waves through the loop.}
\label{xt} 
\end{figure}

In real observations we find a range of frequencies which are present simultaneously in the coronal loops. To mimic this situation, we choose four periods (3,5,7 and 10 minutes) which are being generated all together by a driver acting at one of the loop footpoints. Keeping the wavelength for the longest period in our simulation (which is 100 Mm for the 10 minute period), we choose a loop length of 300 Mm as shown in the upper panel in Figure \ref{xt}. The first and last $\approx$ 18 Mm is left out of the analysis to avoid boundary effects. A rectangular box along the length 
%%%%%%%%%%%%%%%%%%%%%%%%%%%%%%%%%%%%%%%%%%%%%%%%%%%%%%%%%%%%%%%%
%%%%%%%%%%%%%%%%%%%%%%%%%%%%%%%%%%%%%%%%%%%%%%%%%%%%%%%%%%%%%%%%
of the loop, shown in black in upper panel of Figure \ref{xt}, is chosen to construct the time distance plot as before. The time-distance maps for two AIA channel 171 \r{A} and 193 \r{A} for a line of sight (LOS) angle $\theta~=90$\textdegree~ is shown in bottom panels in Figure~\ref{xt}. From these maps, we see different periodicities in the appearances of the inclined ridges in both the channels. We also notice that in these maps, faint reflections generated from the other footpoint are also present. The amplitudes of these reflections are less than 1\% of the input wave amplitudes. Hence, it has negligible effect in the results we produce using these maps.

%%%%%%%%%%%%%%%%%%%%%%%%%%%%%%%%%%%%%%%%%%%%%%%%%%%%%%%%%%%%%%%%
\begin{figure}[h]
\centering
\includegraphics[angle=90,width=0.8\textwidth]{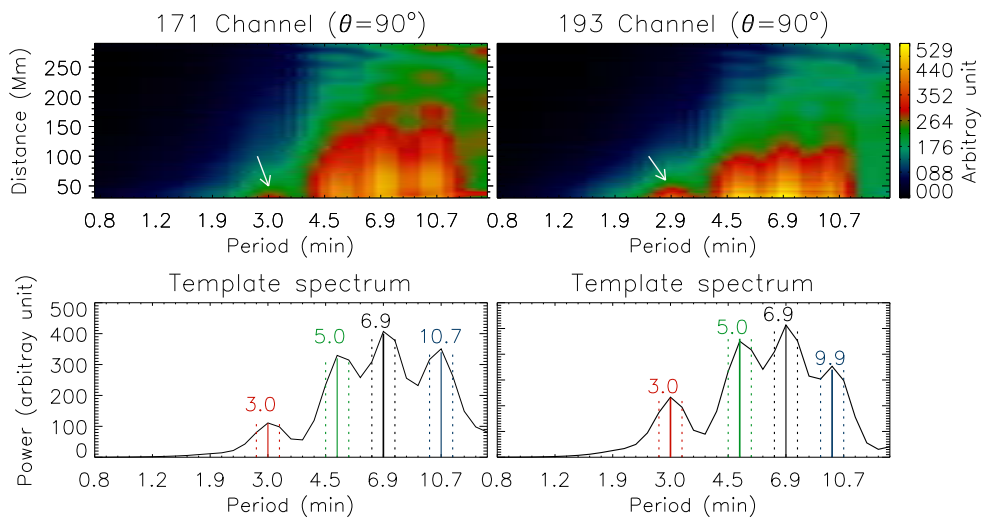}
\caption{$Top:$ The power-distance maps generated from the time-distance maps of the two AIA channels (171~\r{A} and 193~\r{A}). The white arrow in each panel points to the 3 minute power. $Bottom:$ Template spectrum for AIA 193~\r{A} and AIA 171~\r{A} channels respectively. Detected periods (solid lines) are printed on the panels along with their widths (dotted lines) obtained by using $gt\_peaks$.pro.}
\label{power} 
\end{figure}
%%%%%%%%%%%%%%%%%%%%%%%%%%%%%%%%%%%%%%%%%%%%%%%%%%%%%%%%%%%%%%%%

\begin{figure}[h]
\centering
\includegraphics[angle=90,width=1.0\textwidth]{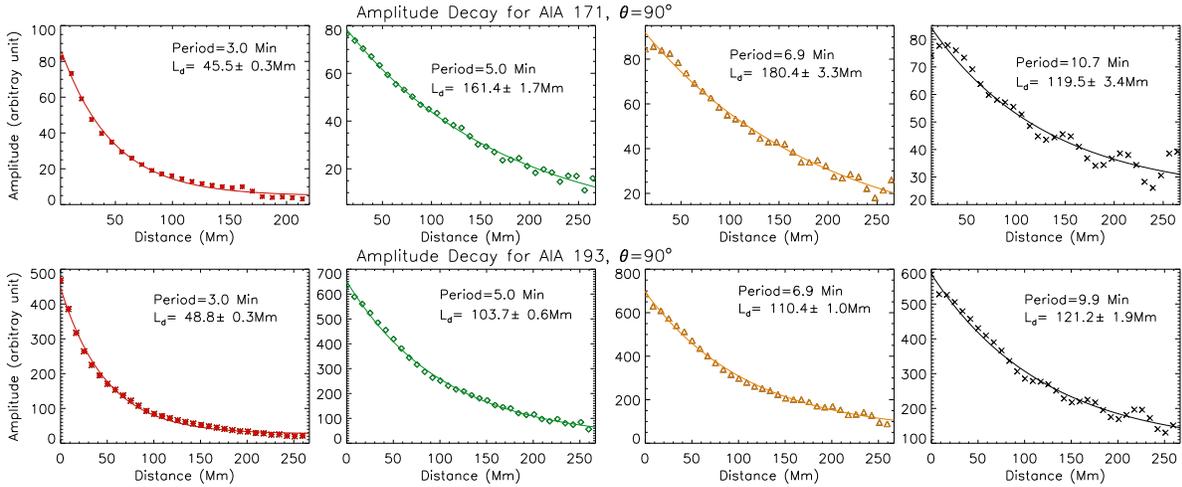}
\caption{$Top:$ Amplitude decay (symbols) of the detected periods for the AIA 171~\r{A} channel for $\theta$=90\textdegree. The fitted exponential function is overplotted as the solid lines. The damping length obtained from the fitting, along with the errors, is printed in each panel. $Bottom:$ Same as previous but for the AIA 193~\r{A} channel for $\theta$=90\textdegree.}
\label{damping} 
\end{figure}

In Figure \ref{power} we show the power-distance maps (upper panels) and their corresponding template spectrum for the 171 \r{A} and 193 \r{A} channel. These power-distance maps act as a tool to separate out the powers confined in different periods and their decay with the distance along the loop. Four distinct periods can clearly be identified from the template spectrum. The slight departure of the detected periods from the given period values in the simulation, can be attributed to the period resolution of the wavelet analysis. All the peaks and their widths in the template spectrum have been identified by the $gt\_peaks$.pro routine. Similar to the single period analysis, we follow the amplitude of a particular period (within the corresponding width) along the distance in the power-distance map. In Figure \ref{damping} we plot the amplitude decay for each detected period for the AIA 171\r{A} and 193\r{A} channels. Here also we plot every alternate $30^{th}$ point to avoid crowding in the amplitude decay plot. The solid lines are the fitted exponential decay function, as described in the previous section and the obtained damping length ($L_d$) from the fitted curve, along with the error, is printed in each panel. We notice from the figure that the damping lengths, in the two AIA channels, are different for the waves with same periods. The difference can be attributed to the different responses of the AIA filters.

%%%%%%%%%%%%%%%%%%%%%%%%%%%%%%%%%%%%%%%%%%%%%%%%%%%%%%%%%%%%%%%%
%%%%%%%%%%%%%%%%%%%%%%%%%%%%%%%%%%%%%%%%%%%%%%%%%%%%%%%%%%%%%%%%

\begin{figure}[h]
\centering
\includegraphics[angle=90,width=0.7\textwidth]{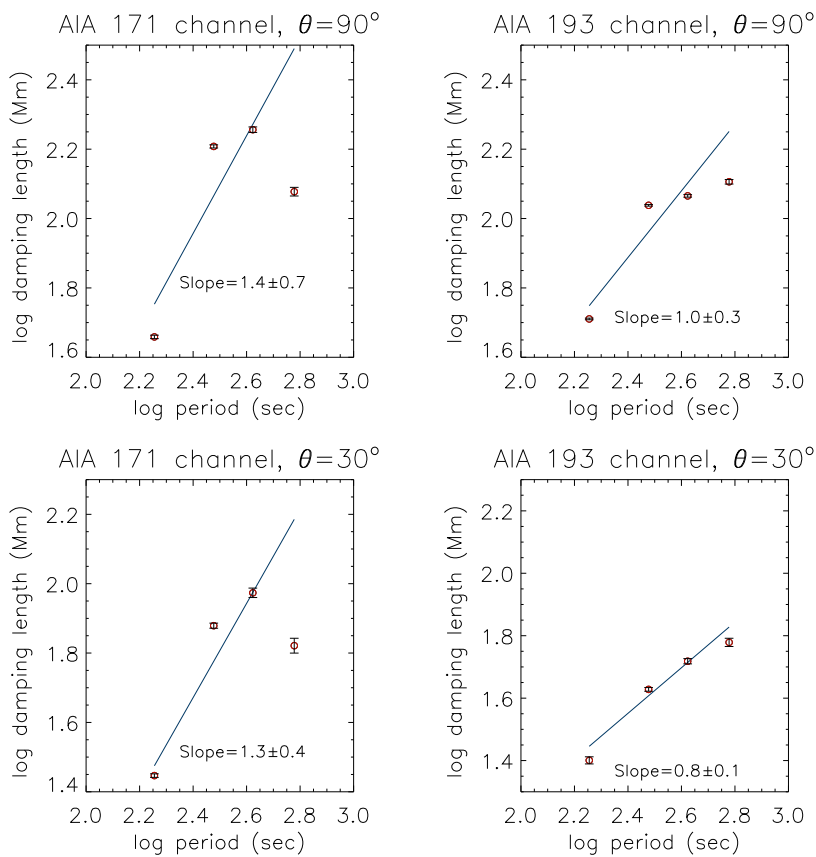}
\caption{The log-log plot of period versus damping length for the two LOS (30\textdegree~and 90\textdegree) for AIA channels 171~\r{A} and 193~\r{A}. The slope of the fitted straight line is printed in each panel along with the error bars.}
\label{log_log} 
\end{figure}

We perform the same analysis for another LOS angle, $\theta =30$\textdegree~for 171 \r{A} and 193 \r{A} channel. The damping lengths obtained from each period, in this case, are listed in Table \ref{table1}. As we are more interested in the change in damping length with the change of period to obtain a frequency/period dependence of the slow waves, we draw a log-log period versus damping length plot (Figure \ref{log_log}) for two AIA channels for two LOS angles.

%%%%%%%%%%%%%%%%%%%%%%%%%%%%%%%%%%%%%%%%%%%%%%%%%%%%%%%%%%%%%%%%%%%%%%%%%%%%%%%%%%
%%%%%%%%%%%%%%%%%%%%%%%%%%%%%%%%%%%%%%%%%%%%%%%%%%%%%%%%%%%%%%%%%%%%%%%%%%%%%%%
Each panel in Figure~\ref{log_log} shows the obtained damping length and the corresponding period in logarithmic scales. Slope values, obtained by fitting a linear function to the data, are printed in each panel. We find the slopes to be positive in all the cases and its value ranges from $\approx +0.8$ to $+1.4$ (with errors less than 0.7). These values match well with the values \citet{2014ApJ...789..118K} found for the sunspot loops. It is worth mentioning that the parameters we have used in our simulation mostly mimic the coronal sunspot loops rather than the polar plumes and interplume regions which have different physical conditions than the sunspot loops.

%New addition
\section{ Analysis of density evolution}

\begin{figure}[h]
\centering
\includegraphics[angle=90,width=0.9\textwidth]{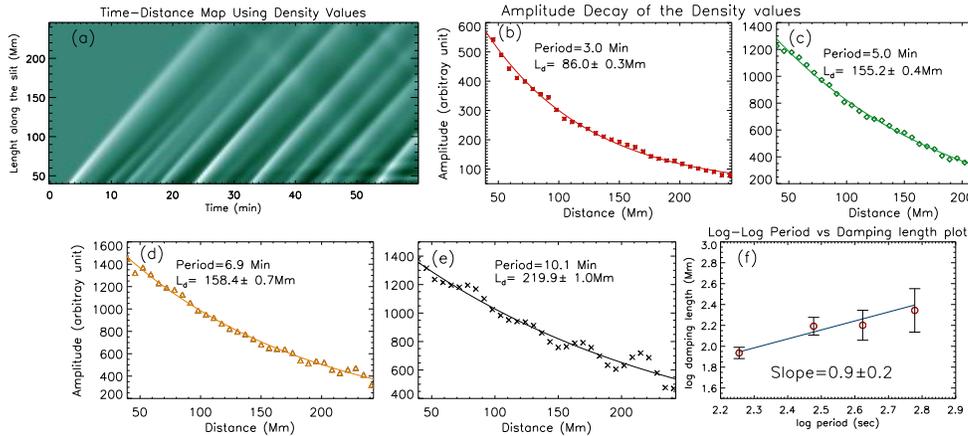}
\caption{ $(a):$ Time-distance map from the density values created using an artificial slit located at loop center, $(b)-(e):$ Amplitude decay for ecah detected period along with the fitted function. Obtained damping lengths are printed in each panel, $(f):$ The log-log period versus damping length plot. Obtained slope (along with the error) is printed in the panel.}
%\caption{Different panels (from top left to bottom right) showing the time-distance map, the damping length curves, using the density values, for different periods and the log-log period versus damping plot respectively. Obtained damping lengths (and the slope) with the errors are printed in respective panels.}
\label{log_log2} 
\end{figure}

\citet{2008SoPh..252..101D} showed that the observed intensity perturbation may not necessarily follow the model density and temperature. To investigate this behaviour, we used the density values, from the simulation and repeated the analysis as above to obtain the log-log period versus damping length plot ( Fig.~\ref{log_log2}). The panel $(a)$ in Figure~\ref{log_log2} shows the time-distance map from the density values obtained by placing an artificial slit along (x=0,y=0), i.e. the centre of the loop. The damping lengths obtained are 86, 155, 158 and 219 Mm (with errors within 1 Mm) for the periods 3, 5, 7, and 10 minutes respectively. Using these damping lengths we obtain a log-log plot of the period versus damping length which has a slope of 0.9$\pm$0.2 . This value is very much consistent with the slope obtained using intensity values. We want to remind the reader that these density values are from the simulation (before the use of FoMo code) and the consistency of this result with the result using the intensity values shows that the emission details are not so important for analyzing the power law begaviour for these waves.

%%%%%%%%%%%%%%%%%%%%%%%%%%%%%%%%%%%%%%%%%%%%%%%%%%%%%%%%%%%%%%%%%%%%%%%%%%%%%%%%%%
%%%%%%%%%%%%%%%%%%%%%%%%%%%%%%%%%%%%%%%%%%%%%%%%%%%%%%%%%%%%%%%%%%%%%%%%%%%%%%%
%upto this

\section{Theory} \label{Theory}

In this study, we consider only the thermal conduction as the slow wave amplitude damping mechanism. Applying a perturbation of the form exp[$i(\omega t - kz) $] on the linearized MHD equations we get the following dispersion relation for the slow waves \citep{2014ApJ...789..118K}

\begin{equation}
   \omega^3 - i\gamma dk^2\omega^2c^2_s - \omega k^2c^2_s + idk^4c^4_s = 0 , 
\label{dispersion}
\end{equation}

where $\omega$ is the angular frequency, \(\gamma\) is the adiabatic index and $c_s$ is the adiabatic sound speed and d the thermal conduction parameter defined as $d$ =~\( \dfrac{(\gamma-1)\kappa_\|T_0}{\gamma c^2_s p_0} \), which has contributions from the equilibrium values of pressure (\(p_0\)), density(\(\rho_0\)), temperature (\(T_0\)) and also from the conductivity (\(\kappa_\|\)) which is parallel to the magnetic field. The damping length ($L_d$) is derived as the reciprocal of the imaginary part of the wave number $k$.

\begin{figure}[htbp]
\centering
\includegraphics[angle=90,width=0.5\textwidth]{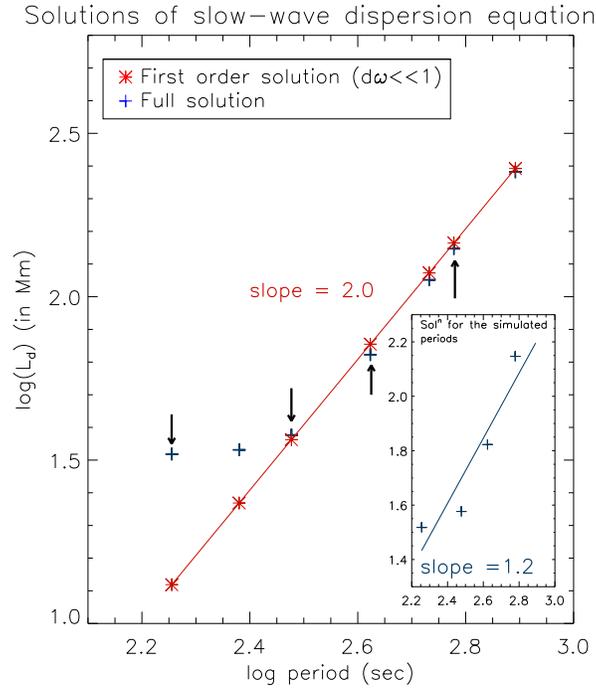}
\caption{Log-log plot of damping lengths obtained by solving the Equation \ref{dispersion} for a range of periods (3 to 13 minutes). The red ($\ast$) represent first order solutions corresponding to the lower thermal conduction limit (d $\omega \ll1$) whereas the blue ($+$) symbols represent the full solutions. The obtained slope for the first order solutions (represented by a red line) is equal to 2 whereas the slope for the full solutions for the frequencies used in the simulation, is equal to 1.2 (as shown in the blue line in the inset plot).}
\label{theory} 
\end{figure}

Equation~\ref{dispersion} is a bi-quadratic equation for k. We have thus solved it analytically for k$^2$ and only retained the solution with the minus sign (corresponding to the solution in Eq.~\ref{dispersion2}). Then k was computed as the square root of the complex k$^2$ and the imaginary part of this was taken as the reciprocal of the damping length $L_d$. To obtain the frequency dependence of L$_d$ we solved the equation for different periods (within the range 3 to 13 minutes) with the parameters we have used in the simulation and plotted the log-log plot in Figure \ref{theory} ($+$ symbols). 

  We also consider the lower thermal conduction limit ($\text{d}\omega \ll1$)) of the Eq.~\ref{dispersion} and the equation reduces to 

\begin{equation}
       k= \dfrac{\omega}{c_s} - i \dfrac{d\omega^2}{2c_s}(\gamma-1)
\label{dispersion2}
\end{equation}
The damping length, under this assumption, is \(\propto 1/\omega^2 \). 
Solutions of the above equation (Eq.\ref{dispersion2}) for periods ranging from 3 to 13 minutes are shown by ($\ast$) symbols in Figure \ref{theory}. The fitted straight line to these points is shown as a red solid line in Figure \ref{theory}.

 From the figure we see that the full and the lower thermal condition limit approximated solutions follow each other very closely for higher periods. For the periods below 5 minutes, the two solutions deviate from each other significantly resulting in a slope less than 2. The d$\omega$ values, for the periods used in our simulation (3,5,7,10 minutes), are 1.07, 0.64, 0.45, 0.32 respectively. So we see that the `small thermal conduction limit' (d$\omega \ll1$) is not valid for smaller periods.

 We also highlight the periods we used for our simulation by the black arrows. To get the power dependence from the theory we fit the full solutions only for the periods used in the simulation and fitted a straight line to these points as shown in the inset in Figure \ref{theory}. The slope measured is equal to 1.2 which is very close to our measured mean value 1.1 .

\section{Summary and Conclusion} \label{Conclusion}

In this study, we set out to explain the observed dependence of the damping of slow waves with their period. In the previous work, no theoretical prediction for damping dependence could explain the observed values. Here we aimed to simulate the damping, perform forward modelling and analyze the model output as in the observations in order to explain the unexpected damping length behaviour with the wave period. 

In our current  study, we first simulated a long coronal loop with a singe period continuous driver at the bottom of the loop footpoint to generate the slow waves. We allowed the waves to decay by applying thermal conduction along the length of the loop. We followed the amplitude of the period along the loop length and fitted an exponentially decaying function to obtain the damping length. In the next phase, we replaced the single period driver with a continuous
 multiperiod driver.

%%%%%%%%%%%%%%%%%%%%%%%%%%%%%%%%%%%%%%%%%%%%%%%%%%%%%%%%%%%%%%%%%%%%%%%%%%%%%%%%%%%%%%%%%%%%
%%%%%%%%%%%%%%%%%%%%%%%%%%%%%%%%%%%%%%%%%%%%%%%%%%%%%%%%%%%%%%%%%%%%%%%%%%%%%%%%%%%%%%%%%%%

 Following the same approach as above, we drew a log-log plot of the period versus damping length and fit a linear function to obtain a power law index. We obtained a slope of 1.4$\pm$0.7 (171\r{A} channel) and 1.0$\pm$0.3 (193\r{A} channel) for \(\theta~=90$\textdegree $\) and 1.3$\pm$0.4 (171\r{A} channel) and 0.8$\pm$0.1 (193\r{A} channel) for \(\theta~=30$\textdegree$ \). We conclude from these values that the wave is getting damped linearly with change of period. Furthermore, to see the effect of the emissivity and the instrumental responses on the power law index, we have used the density values obtained from the simulation. The identical slope values indicate that these factors are not important in these studies.

We have also studied in-depth the theoretical damping behaviour of the slow waves in the case when thermal conduction is not weak. Using linear theory and including thermal conduction as the dominant damping mechanism, one would obtain a slope of 2 in a log-log period versus damping length plot \citep{2014ApJ...789..118K}. We have found that in the general case (and not the weak conduction limit previously used) a similar positive slope, as in the observation, can be reproduced. This shows that the 'lower thermal conduction limit' (d$\omega\ll1$) assumption is not valid for lower periods which in terms shifts the slope towards lower positive value than the theoretically predicted value of 2.

 On the other hand the negative slopes for polar plume and the interplume case obtained by \citet{2014ApJ...789..118K} are still to be explained. They are possibly due to different nature of the polar plumes compared to the sunspot loops. The density and temperature structures in plumes are different from sunspot loops and change very rapidly with the height from the plume footpoint \citep{2006A&A...455..697W}. We have not considered any magnetic field divergence, density stratification in our loop model, and thus the negative slopes may indicate a different dominant damping source other than the thermal conduction.

\section{Acknowledgment}

This research has been made possible by the topping-up grant CHARM+top-up COR-SEIS of the BELSPO and the Indian DST. It was also sponsored by an Odysseus grant of the FWO Vlaanderen, and Belspo's IUAP CHARM.
%---------------------E-N-D---------------------------------

%%%%%%%%%%%%%%%%%%%%%%%%%%%%%%%%%%%%%%%%%%%%%%%%%%%%%%%%%%%%%%%%%%%%%%%%%%%%%%%%%%%%%%%%%%%%%%%%%%%%%%%%%%%%%%%%%
 \begin{table}[H]
\begin{center}
\caption{Damping Lengths obtained for \(~\theta~= 30$\textdegree $\)}  

\label{table1}
\begin{tabular}{lcccc r@{   }l c} % define the column alignment
                                  % l: left, c: center, r: right
                                  % @{.} replace the inter-column by a .
  \hline
  & \multicolumn{1}{c}{$AIA$ } & \multicolumn{1}{c}{$Period$} & \multicolumn{1}{c}{$Damping~Length$} \\
&$ Channel $  & $(Min)$ & $(Mm)$ \\

     \hline
     & 171~\r{A} & 3.0 & 27.9$\pm$ 0.3 \\
     &           & 5.0 & 75.6$\pm$ 0.7\\
     & ~~        & 6.9 & 94.1$\pm$ 1.2\\
     & ~~        & 10.7& 66.2$\pm$1.3\\
\hline
     & 193~\r{A} &  3.0 & 25.1$\pm$ 0.3\\
     &           &  5.0 & 42.5$\pm$ 0.5\\
     &           &  6.9 & 52.3$\pm$ 0.5\\
     &           &  9.8 & 60.1$\pm$ 0.7\\
  \hline

\end{tabular}
\end{center}
\end{table}
% 
% %%%%%%%%%%%%%%%%%%%%%%%%%%%%%%%%%%%%%%%%%%%%%%%%%%%%%%%%%%%%%%%%%%%%%%%%%%%%%%%%%%%%%%%%%%%%%%%%%%%%%%%%%%%%%%%%%%%%

%%%--------------------------------------------------------------------------------------------------------------

\textbf{Appendix}

The dispersion relation for the slow wave from the linearized MHD equations, including the thermal conduction, reads as:

\begin{align}
   \omega^3 - i\gamma dk^2\omega^2c^2_s - \omega k^2c^2_s + idk^4c^4_s = 0 , 
\label{appendix_eq}
\end{align}

where $\omega$ is the angular frequency, \(\gamma\) is the adiabatic index and $c_s$ is the adiabatic sound speed and d the thermal conduction parameter defined as $d$ =~\( \dfrac{(\gamma-1)\kappa_\|T_0}{\gamma c^2_s p_0} \), which has contributions from the equilibrium values of pressure (\(p_0\)), density(\(\rho_0\)), temperature (\(T_0\)) and also from the conductivity (\(\kappa_\|\)) which is parallel to the magnetic field.

First we consider the simplest case when $d=0$ which gives:
  \begin{align*}
     k= \frac{\omega}{c_s} 
  \end{align*}

Now we consider the case when d$\omega\ll1$ and we assume the solution to be in the form 
\begin{align*}
     k= \frac{\omega}{c_s}+d{\omega_1} , 
\end{align*}
where $\omega_1$, the first order coefficient of d, is the correction term for this case.

Putting this value of k in Eq. \ref{appendix_eq} we get

\begin{align*}
   \omega^3 - i\gamma d(\frac{\omega}{c_s}+d{\omega_1})^2\omega^2c^2_s - \omega (\frac{\omega}{c_s}+d{\omega_1})^2c^2_s + id(\frac{\omega}{c_s}+d{\omega_1})^4c^4_s = 0 ,
\end{align*}

Expanding the terms and considering the coefficients of the first order terms in d yields,
%\begin{equation*}
%\begin{multiline}
\centering
\( \Longrightarrow - i\gamma \omega^4  -2\omega^2 \omega_1 c_s +i \omega^4=0\)\\
\( \Longrightarrow(\gamma-1)\omega^4 -2i\omega^2\omega_1 c_s=0 ~~~ \)\\
\(\Longrightarrow \omega_1= -\frac{i\omega^2(\gamma-1)}{2c_s}\)
%\end{multiline}
%\end{equation*}

Thus the solution at the lower thermal conduction limit, becomes

\begin{equation*}
k= \frac{\omega}{c_s} -\frac{id\omega^2(\gamma-1)}{2c_s}
\end{equation*}

 \bibliographystyle{apj}
 \bibliography{references}

\end{document}